\documentclass[twocolumn,pra,superscriptaddress,floatfix,longbibliography]{revtex4-2}
\usepackage{graphicx}
\usepackage{xcolor}
\usepackage{hyperref}
\usepackage{braket}
\usepackage{bm}
\usepackage{bbold}
\usepackage{soul}

\expandafter\let\csname equation*\endcsname\relax 
\expandafter\let\csname endequation*\endcsname\relax
\usepackage{amsmath}
\usepackage{cleveref}

\usepackage{ulem}  
\newcommand{\ssection}[1]{{\noi  \it #1:}}
\newcommand{\ketbra}[2]{|\,{#1}\,\rangle \langle\,{#2}\, |}

\def\noi{\noindent}
\def\beq{\begin{equation}}
	\def\eeq{\end{equation}}

\usepackage{ulem}  
\usepackage{color}

\begin{document}
	
	\title{Quantum simulation of Motzkin spin chain with Rydberg atoms}


	\author{Kaustav Mukherjee}
	\affiliation{Department of Physics and Astronomy, University of Tennessee, Chattanooga, TN 37403, USA}
	\affiliation{UTC Quantum Center, University of Tennessee, Chattanooga, TN 37403, USA}	\email{Kaustav-Mukherjee@utc.edu}

		\author{Hatem Barghathi}
		\affiliation{Department of Physics and Astronomy, University of Tennessee, Knoxville, Tennessee 37996, USA}
		\affiliation{Institute for Advanced Materials and Manufacturing, University of Tennessee, Knoxville, Tennessee 37996, USA}
		
				\author{Adrian Del Maestro}
		\affiliation{Department of Physics and Astronomy, University of Tennessee, Knoxville, Tennessee 37996, USA}
		\affiliation{Institute for Advanced Materials and Manufacturing, University of Tennessee, Knoxville, Tennessee 37996, USA}
        \affiliation{Min H. Kao Department of Electrical Engineering and Computer Science, University of Tennessee,
Knoxville, Tennessee 37996, USA}
		
			\author{Rick Mukherjee}
					\affiliation{Department of Physics and Astronomy, University of Tennessee, Chattanooga, TN 37403, USA}
					\affiliation{UTC Quantum Center, University of Tennessee, Chattanooga, TN 37403, USA}

	\vspace{10pt}

	\begin{abstract}
Motzkin spin chain is a well-known mathematical model with connections to symmetry-protected topological phases, such as the Haldane phase, as well as to concepts in the AdS/CFT correspondence.
They exhibit highly entangled ground states that violate the area law and are exceptionally difficult to simulate with conventional numerical methods. 
Numerical simulations of the Motzkin ground state become further challenging at large system sizes due to their high-dimensional spin structure, rendering it a natural test bed for quantum simulation with ultra-cold systems. Here, we propose a Rydberg-atom based quantum simulation scheme that effectively realizes Motzkin spins using an experimentally accessible set of parameters. We show that the resulting effective Motzkin ground state reproduces the characteristic entanglement scaling and the block-structure properties of the reduced density matrix associated with the ideal Motzkin state. Our results establish a pathway toward a concrete experimental realization of Motzkin spins beyond purely mathematical constructions, opening avenues for exploring other similar exotic non–area-law entangled phases in programmable Rydberg simulators.
	\end{abstract}
	
	\maketitle
	
	\ssection{Introduction}
Ground states that violate the area law of entanglement are very relevant as they provide rare examples of highly entangled phases that lie beyond the reach of conventional numerical tools. Notable physical models exhibiting such entanglement include fine-tuned free-fermion models \cite{gioev2006entanglement,wolf2006violation,swingle2012conformal,gori2015explicit}, bosonic systems \cite{lai2013violation,ding2009entanglement}, interacting spin chains \cite{eisert2006general,its2005entanglement,popkov2005logarithmic,ding2008block,latorre2009short,koffel2012entanglement,buyskikh2016entanglement,frerot2017entanglement,vodola2015long} and certain gapless one-dimensional systems \cite{calabrese2009entanglement}. One particularly interesting example is the Motzkin spin chain \cite{bravyi2012criticality,movassagh2016supercritical}, a mathematically constructed model whose ground state violates the area law despite being governed by a frustration-free Hamiltonian with a spectral gap that closes algebraically with system size. The violation of the area law stems from an underlying conservation law that constrains the accessible Hilbert space, thereby circumventing the assumptions of Hastings’ theorem \cite{hastings2007area}.

Despite the remarkable properties of the Motzkin spin model, it has largely remained a theoretical construct. 
Classical simulation techniques, including matrix-product-state (MPS) methods \cite{schuch2008entropy} such as DMRG \cite{white1992density,schollwock2005density}, and tensor-network approaches like projected entangled pair states (PEPS) \cite{orus2014practical}, can simulate systems with weak (logarithmic) violations of the area law. However, the applicability of these classical techniques becomes restrictive for systems with higher-dimensional spins and strong area-law violation. Both of these features exist with the Motzkin spin model \cite{bravyi2012criticality,movassagh2016supercritical}, hence making it a compelling case for quantum simulation \cite{blatt2012quantum,monroe2021programmable,weimer2010rydberg,georgescu2014quantum}. 
The realization of Motzkin spin chains is of broad interest, with potential applications involving
exploration of symmetry-protected topological order such as the Haldane phase \cite{barbiero2017haldane}, and connections to the AdS/CFT correspondence \cite{alexander2019exact,alexander2021exact}.

In this work, we propose a Rydberg atom-based quantum simulation platform \cite{Adams2020,scholl2021quantum,browaeys2020many,bernien2017probing,Notarnicola_2023} that effectively implements the local constraints of the Motzkin spin model. Leveraging the 
state-dependent interactions \cite{younge2010state} and dipolar couplings \cite{gallagher2008dipole,Browaeys2016}, we establish a Motzkin-state preparation protocol, that
provides a route toward engineering an effective Motzkin ground state within state-of-the-art experiments \cite{bernien2017probing,Notarnicola_2023,Guillaume2024,manetsch2025tweezer}. Our framework not only establishes a route towards the physical realization of Motzkin spins, but also provides a platform for systematically probing their distinctive entanglement properties and many-body dynamics.

While Rydberg interactions alone cannot realize the Motzkin spin model, we additionally use local detunings to prepare an effective ground state with high fidelity 
with respect to the ideal Motzkin ground state for small system sizes. Furthermore, we investigate the entanglement scaling and the structure of the reduced density matrix of the final state obtained with our protocol, and find agreement with those of the ideal Motzkin ground state. We show that the effective ground state reproduces the logarithmic violation of area-law entanglement scaling characteristic of the original Motzkin model.

    \begin{figure}
        \centering
        \includegraphics[width=0.99\linewidth]{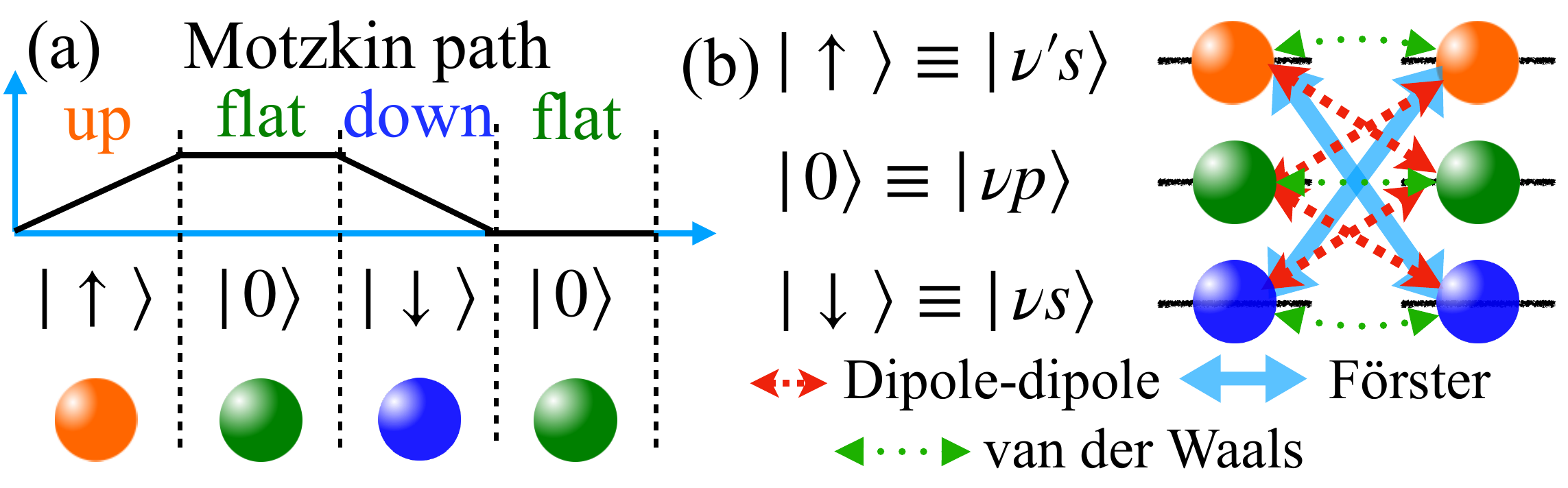}
        \caption{(a) An example of a Motzkin path with respective encoding of each path with the three levels of spin-1 particle, denoted by $\{\ket{\uparrow},\ket{0},\ket{\downarrow}\}$. (b) Encoding the three levels of spin-1 particle into Rydberg atoms with principal quantum numbers $\nu$ and $\nu'$, and angular quantum numbers $s$ $(l=0)$ and $p$ $(l=1)$. Red-dashed lines and green-dotted arrows indicate spin-exchange dipole-dipole interactions and van der Waals based energy shifts, respectively, while the blue arrows indicate F\"orster resonance interactions.}
        \label{fig:schematic}
    \end{figure}
\ssection{Theory} 
Motzkin spin chains consist of an array of qudits and appear in two primary variants: the \emph{colorless} version \cite{bravyi2012criticality}, constructed from qutrits (spin-1 particles with three internal levels), and the \emph{colored} version \cite{movassagh2016supercritical}, built from spin-2 particles with five internal levels. The colorless Motzkin model already exhibits logarithmic violations of the area-law entanglement scaling, whereas the colored variant exhibits power-law scaling despite being gapped, frustration-free and local. In this work, we present a proof-of-principle demonstration of realizing the colorless spin-1 Motzkin model in a Rydberg-atom quantum simulator. Using the same principle, our protocol can be extended to colored Motzkin spins by including additional internal levels and couplings. 

 The three levels of each spin-1 site are denoted by $\{\ket{\uparrow},\ket{0},\ket{\downarrow}\}$ and are mapped to the three allowed steps of a Motzkin path: ``up'', ``flat'', and ``down'', respectively as shown in Fig.~\ref{fig:schematic}~(a) \cite{bravyi2012criticality}. A valid Motzkin path of length $N$ is a sequence 
\[
\ket{\boldsymbol{\sigma}} = \ket{\sigma_1 \sigma_2 \cdots \sigma_N}, \qquad 
\sigma_i \in \{\uparrow,0,\downarrow\},
\] 
that satisfies three constraints: (i) it begins at height $0$, (ii) ends at height $0$, and (iii) never goes below the horizontal axis at intermediate positions. These set of constraints ensures that for every $\ket{\uparrow}$ is accompanied by a corresponding $\ket{\downarrow}$. 
The Motzkin ground state is an equal superposition of all such valid paths,
\begin{equation}
    \ket{\psi_{\rm Motzkin}}
    = \frac{1}{\mathcal{N}} 
    \sum_{\text{Motzkin paths } \boldsymbol{\sigma}} 
    \ket{\boldsymbol{\sigma}},
    \label{Motzkin_state}
\end{equation}
where $\mathcal{N}$ is the normalization constant. Examples of Motzkin ground state for small systems, such as $N=2$ and $N=3$ are $\ket{\psi}_2 =  \frac{1}{\sqrt{2}}(\ket{\uparrow\downarrow}+\ket{00})$ and $\ket{\psi}_3 =  \frac{1}{2}(\ket{\uparrow\downarrow0}+\ket{\uparrow0\downarrow}+\ket{0\uparrow\downarrow}+\ket{000})$, respectively.

The simplest Hamiltonian admitting this state as its \emph{unique} ground state is a frustration-free, nearest-neighbor model of the form
\begin{equation}
    H =  \frac{1}{2}\sum_{i=1}^{N-1}\Pi_{i,i+1} + \Pi_{\text{boundary}} ,
    \label{full_Motzkin}
\end{equation}
where the projectors $\Pi_{i,i+1}$ act on $i$th and $(i+1)$-th site of the system as,
\begin{align}
    \Pi_{i,i+1} &= - \bigg(\ketbra{\downarrow0}{0\downarrow} +\ketbra{\uparrow0}{0\uparrow} + \ketbra{\uparrow\downarrow}{00} + \text{h.c.}\bigg) \nonumber\\
    &+
   \bigg\{ \ketbra{\uparrow0}{\uparrow0} + \ketbra{0\uparrow}{0\uparrow} +
    \ketbra{\downarrow0}{\downarrow0} \nonumber \\
    &+ \ketbra{0\downarrow}{0\downarrow} +  \ketbra{00}{00}\bigg\} +
   \bigg[ \ketbra{\uparrow\downarrow}{\uparrow\downarrow} \bigg]  ,
    \label{Motzkin_Hamil}
\end{align}
and the boundary term 
\begin{equation}
    \Pi_{\text{boundary}}
    = \ket{\downarrow}_1\bra{\downarrow} + \ket{\uparrow}_N\bra{\uparrow}
    \label{Motzkin_boundary}
\end{equation}
penalizes configurations that violate path-height constraints at the edges, such as a down-step at the first site and an up-step at the last site. The operator $\ket{}_n\bra{}$ acts on the $n$th site of the chain. We decompose the Motzkin Hamiltonian into three groups, distinguished by circular, curly, and square brackets as shown in Eq.~\ref{Motzkin_Hamil}. 
More details about the Motzkin spin chain Hamiltonian are provided in Refs.~\cite{bravyi2012criticality,movassagh2016supercritical,SI}. 

\ssection{Implementation of Motzkin Hamiltonian in Rydberg simulator}
Rydberg atoms are neutral atoms with large principal quantum numbers $\nu\gtrsim 10$ \cite{gallagher2006rydberg}, providing a versatile and highly tunable platform for quantum simulation owing to their long-range dipole-dipole interactions \cite{gallagher2008dipole,Browaeys2016}, strong van der Waals couplings \cite{PhysRevLett.110.263201}, long coherence times \cite{minns2006preserving,levine2018high} and deterministic positioning with optical tweezers \cite{barredo2018synthetic}. Their unique properties have enabled quantum simulations of diverse condensed-matter systems \cite{scholl2021quantum,mukherjee2024excitons}, high-energy physics Hamiltonians \cite{vovrosh2022dynamical,wang2024quantum}, exotic ground state phases \cite{zeybek2025formation,zeybek2024bond}, frustrated spin models \cite{glaetzle2015designing,tian2025engineering} and  disordered spin systems \cite{signoles2021glassy,mukherjee2024influence}.

We consider a chain of $N$ atoms with three relevant Rydberg levels 
$\{\ket{\nu s},\, \ket{\nu p},\, \ket{\nu' s}\} \equiv \{\ket{\uparrow},\ket{0},\ket{\downarrow}\}$,
which form an effective spin-1 system. The dominant interactions include dipole-dipole exchange \cite{gallagher2008dipole,Browaeys2016} (shown as red-dashed arrows in Fig.~\ref{fig:schematic}~(b)), van der Waals shifts \cite{PhysRevLett.110.263201} (green-dotted arrows), and Förster resonances \cite{ryabtsev2010observation,ravets2014coherent,beterov2015rydberg,mogerle2025spin} (blue-solid arrows), altogether leading to the Hamiltonian
\begin{equation}
    \hat{H} = 
    \hat{H}_{\mathrm{dip}}
    + \hat{H}_{\mathrm{vdW}}
    + \hat{H}_{\mathrm{Forster}},
    \label{full_Hamil}
\end{equation}
where the dipole-dipole Hamiltonian is written as
\begin{align}
    \hat{H}_{\mathrm{dip}} 
    = \sum_{i\neq j} 
    & J^{\uparrow0}_{ij} \big(\ket{ \uparrow 0}_{ij}\bra{0\uparrow}+ \text{h.c.}\big) \nonumber\\
    +\;& J^{0\downarrow}_{ij} \big(\ket{\downarrow 0}_{ij}\bra{0\downarrow} + \text{h.c.}\big)\nonumber\\
    +\;& J^{00}_{ij}\big( \ket{\uparrow\downarrow}_{ij}\bra{00}    + \ket{\downarrow \uparrow }_{ij}\bra{00}  
        + \text{h.c.} \big),
    \label{dipole_1}
\end{align}
with $J^{\alpha\beta}_{ij}=C_3^{\alpha\beta}(1-3\cos^2\theta_{ij})/a_{ij}^3$ and $\ket{}_{ij}\bra{}$ denotes a two-body operator acting on $i$th and $j$th atom. Here $a_{ij}$ is the distance between the $i$th and $j$th atom and $\theta_{ij}$ indicates the relative angle between the pair of atoms with respect to the direction of the external magnetic field.  While the $J^{\uparrow0}_{ij}$ and $J^{0\downarrow}_{ij}$ terms represent conventional spin-exchange interactions, the $J^{00}_{ij}$ term couples the pair states $\ket{\uparrow \downarrow}$ and $\ket{\downarrow \uparrow}$ with $\ket{00}$. The simultaneous exchange process $\ket{\uparrow}\otimes\ket{\downarrow}\leftrightarrow \ket{0}\otimes\ket{0}$ is enabled only under resonant conditions, requiring energy matching between the two involved pairs of neighboring levels \cite{mogerle2025spin}. These dipole-dipole interaction terms in $\hat{H}_{\mathrm{dip}}$ map onto the terms in the circular bracket of Eq.~\ref{Motzkin_Hamil}.

The van der Waals interaction contributes to the energy shifts leading to the diagonal terms,
\begin{align}
    \hat{H}_{\mathrm{vdW}}
    &= \sum_{i\neq j}\sum_{\alpha\beta\in\{\uparrow\uparrow,\downarrow\downarrow,00,\uparrow0,\downarrow0\}} 
    V^{\alpha\beta}_{ij} \ket{\alpha \beta}_{ij}\bra{\alpha \beta} ,
    \label{vdw_Hamil}
\end{align}
where $V^{\alpha\beta}_{ij}=C_6^{\alpha\beta}/a_{ij}^6$, with $\alpha$ and $\beta$ restricted to same or neighboring Rydberg levels. These terms reproduce the expression in the curly bracket of Eq.~\ref{Motzkin_Hamil}.

In addition to dipole-dipole and van der Waals interactions, which naturally couple neighboring Rydberg levels, couplings between non-neighboring Rydberg levels can be induced via F\"orster resonances by tuning nearby Rydberg levels into resonance with the $\{\ket{\uparrow},\ket{0},\ket{\downarrow}\}$ manifold \cite{ryabtsev2010observation,ravets2014coherent,beterov2015rydberg,mogerle2025spin}, resulting in
\begin{align}
    \hat{H}_{\mathrm{Forster}}
    = \sum_{i\neq j}
    V_{ij}^{\mathrm{diag}} \big( \ket{\uparrow\downarrow}_{ij}\bra{\uparrow\downarrow} + \ket{\downarrow \uparrow }_{ij}\bra{\downarrow\uparrow}\big) \nonumber\\ 
    + V_{ij}^{\mathrm{ofd}} \big( \ket{\uparrow\downarrow}_{ij}\bra{\downarrow\uparrow} + \text{h.c.}\big).
    \label{Forster_Hamil}
\end{align}
where $V_{ij}^{\mathrm{diag/ofd}}=9 \sin^2 \theta_{ij} \cos^2\theta_{ij} C_6^{\text{F{\"o}rster}}/a_{ij}^6\Delta$ \cite{mogerle2025spin}, and $\Delta$ is the detuning from the F{\"o}rster resonant level. The F{\"o}rster interactions correspond to the expression in the square bracket of Eq.~\ref{Motzkin_Hamil}.

\begin{figure}
    \centering
    \includegraphics[width=0.99\linewidth]{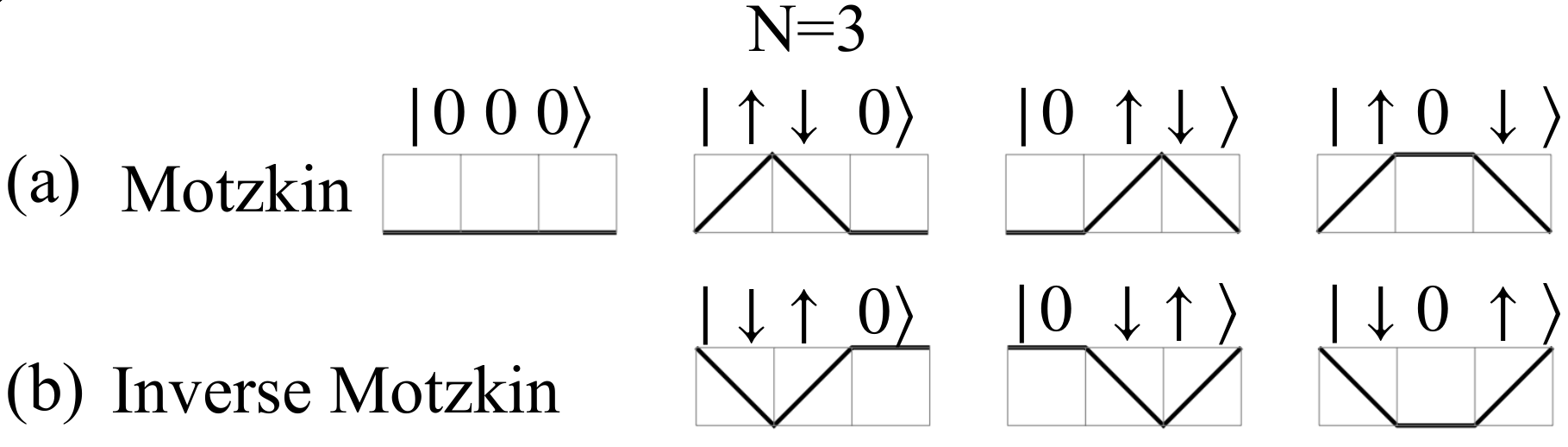}
    \caption{(a) Motzkin and (b) Inverse Motzkin states, for $N=3$, shown in both spin and path representations. Motzkin paths remain non-negative at all steps, whereas inverse-Motzkin paths violate this constraint and form the vertical mirror counterparts of Motzkin paths.}
    \label{fig:inverse_motzkin}
\end{figure}

While the Rydberg Hamiltonian in Eq.~\ref{full_Hamil} captures all the terms of the Motzkin Hamiltonian, it inevitably includes unwanted additional terms. These additional terms are: 
(i) a spin-exchange term $J^{00}_{ij}(\ket{\downarrow \uparrow }_{ij}\bra{00}+\mathrm{h.c.})$, (ii) a diagonal interaction $V^{\mathrm{diag}}_{ij}\ket{\downarrow \uparrow }_{ij}\bra{\downarrow\uparrow}$, and (iii) an off-diagonal coupling $V^{\mathrm{ofd}}_{ij}(\ket{\uparrow\downarrow}_{ij}\bra{\downarrow\uparrow}+\mathrm{h.c.})$. As a consequence, the Rydberg Hamiltonian gives rise to additional states that can be classified as inverse-Motzkin states, shown in Fig.~\ref{fig:inverse_motzkin}~(b), respectively. While the inverse-Motzkin states individually exhibit structural features analogous to conventional Motzkin states, shown in Fig.~\ref{fig:inverse_motzkin}~(a), their simultaneous presence alongside Motzkin states diminishes the characteristic features of the Motzkin spin model. 

To address the additional states, we explore two approaches. 
First, the fine-tuning model, in which the exact Motzkin state emerges as the ground state through a precise cancellation of the additional states achieved by tuning the Rydberg interaction parameters. This requires satisfying the following set of conditions: 
$
J^{00}_{i,i+1} = -V^{00}_{i,i+1} = -V^{\mathrm{diag}}_{i,i+1} = -V^{\mathrm{ofd}}_{i,i+1}, 
\qquad
J^{\downarrow0}_{i,i+1} = V^{\downarrow0}_{i,i+1}, 
\qquad
J^{+0}_{i,i+1} = V^{+0}_{i,i+1}.
$
For more details on energy spectra and experimentally realistic parameters, see section \ref{SI:interaction_para} of SM~\cite{SI}.
 While this approach results in an exact Motzkin ground state, it is highly sensitive to parameter fluctuations, with small deviations sufficient to destroy the Motzkin structure. 

\begin{figure}
    \centering
    \includegraphics[width=0.99\linewidth]{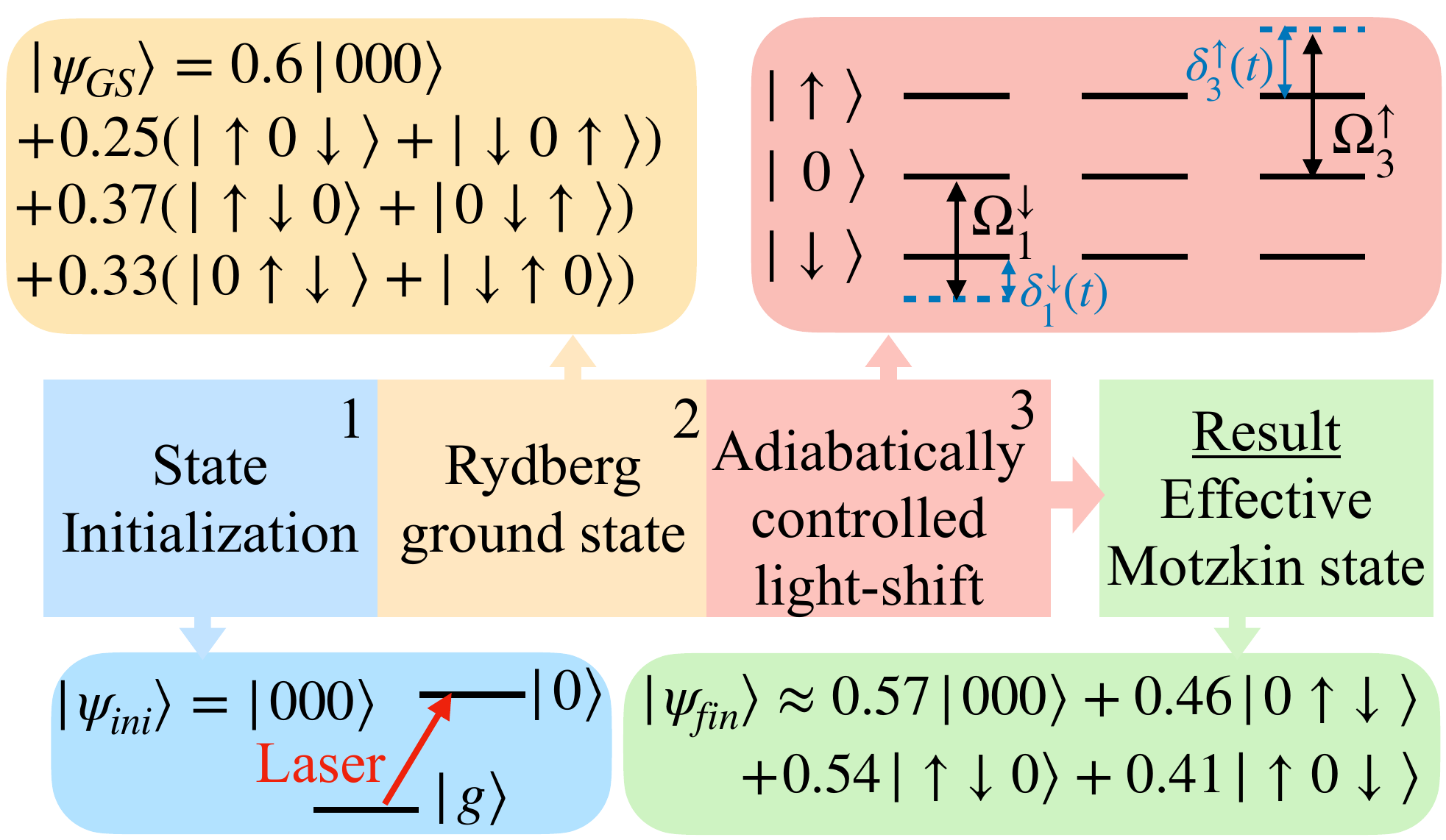}
    \caption{Adiabatic-control protocol: (1) state initialization, (2) preparation of the Rydberg ground state, (3) application of a slowly-varying locally controlled microwave coupling denoted by $\Omega_{1,3}^{\uparrow,\downarrow},\delta_{1,3}^{\uparrow,\downarrow}$  and (4) emergence of the effective Motzkin state.}
    \label{fig:protocol}
\end{figure}

Alternatively, the additional states can be suppressed using an adiabatic control protocol comprising three steps,
as shown in Fig.~\ref{fig:protocol}. First, the Rydberg array is initialized in a product state $\ket{\psi_{\mathrm{ini}}}=\ket{00\cdots0}$, where each atom is excited to the Rydberg level $\ket{0}$. 
Our numerical investigation indicates that, in order to obtain a high-fidelity Motzkin state as the ground state, the system requires to be prepared in its ground state, from which it can be adiabatically evolved into the Motzkin ground state. This initialization step [yellow block in Fig.~\ref{fig:protocol}] involves evolving the system to the ground state via quantum optimal control \cite{khaneja2005optimal,de2011second}, which we discuss in Sec.~\ref{SI:ground_state_prep} of the SM \cite{SI}. 
Preparation of specific target states using quantum optimal control in Rydberg platforms has been demonstrated both theoretically \cite{ostmann2017non,mukherjee2020preparation,kuros2021controlled,PhysRevLett.125.203603} and experimentally \cite{xu2021fast,browaeys2020many,lu2024probing,chen2024quantum}, supporting the experimental feasibility of the initialization step. 
In the final step, the Rydberg ground state is adiabatically evolved into the effective Motzkin ground state using controlled detuning of local microwave couplings \cite{khaneja2005optimal,fuller2025adiabatic,van2023adiabatic,scott2025generalized}. These detunings selectively penalizes specific internal states at chosen sites, for instance, $\ket{\downarrow}$ at site $1$ and $\ket{\uparrow}$ at site $N$, as depicted in the red-block of Fig.~\ref{fig:protocol}, suppressing the inverse-Motzkin states of the form $\ket{\downarrow\cdots\uparrow}$.  The Hamiltonian for the local microwave couplings with controlled detunings is described by
\begin{align}
    \hat{H}_{\mathrm{control}}(t) = &\sum_{i=1}^N \sum_{\alpha\in\{\uparrow,\downarrow\}} \bigg[\frac{\Omega_i^{\alpha }}{2}(\ket{0}_i\bra{\alpha}+h.c.)\nonumber\\
    &+
    \delta_i^\alpha(t)\,\ket{\alpha}_i\bra{\alpha}\bigg],
    \label{control}
\end{align}
where $i$ labels the atomic site, $\Omega_i^{\alpha }$ and $\delta_i^\alpha(t)$ denotes the Rabi frequency and detuning, respectively, of the microwave coupling $\ket{\alpha}\leftrightarrow\ket{0}$, where $\alpha\in\{\uparrow,\downarrow\}$. We fix $\Omega_i^{\alpha }=0.1$ MHz, as it plays a minimal role in the protocol. In contrast, the detunings $\delta_i^\alpha(t)$ constitute the key control parameters in the adiabatic preparation. These detunings are linearly ramped from $0$ to a maximum of $200$ MHz over a protocol duration of $10-20\mu$s, corresponding to a maximum ramp rate of $\dot{\delta}_i^\alpha = 20$ MHz/$\mu$s, which is slower than the minimum interaction energy scale ($\sim 35~\mathrm{MHz}$), thus ensuring adiabaticity.

    \begin{figure}
        \centering
        \includegraphics[width=0.95\linewidth]{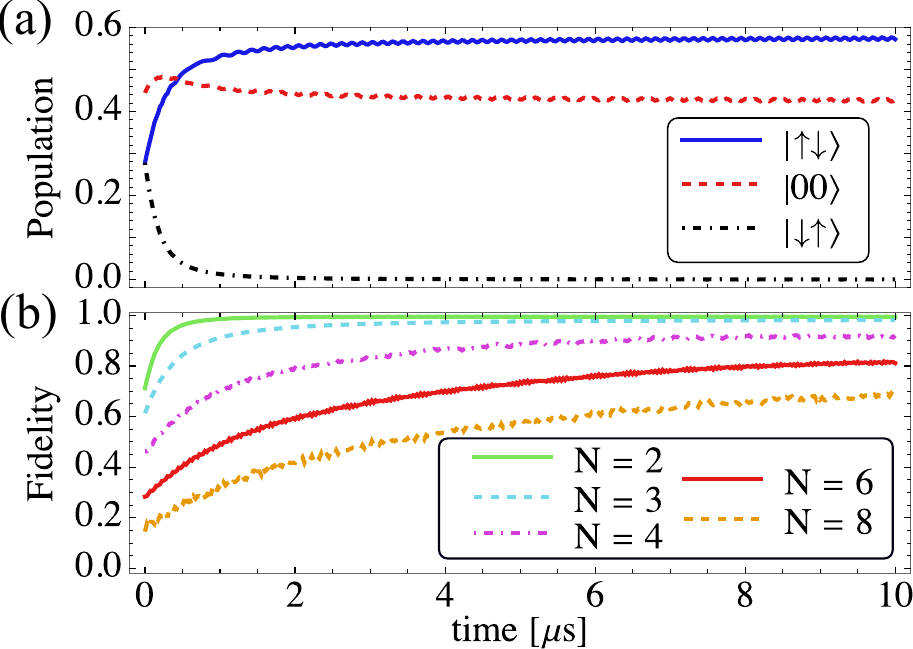}
        \caption{(a) Population dynamics of the $N=2$ Rydberg system, showing the suppression of undesired states and the emergence of the effective Motzkin ground state. (b) Fidelity $\mathcal{F}$ of the effective ground state with respect to the Motzkin ground state for system sizes $N = 2$ to $8$. }
        \label{fig:population_fidelity}
    \end{figure}
    
\ssection{Results}
The results for the adiabatic control protocol are obtained using $^{87}\mathrm{Rb}$ atoms, with the relevant level structure and interaction parameters detailed in Sec.~\ref{SI:interaction_para} of the SM \cite{SI}. As a minimal example, we first consider a system of $N=2$, for which the Rydberg ground state is
$\ket{\psi_{\mathrm{GS}}}=0.67\ket{00}+0.53(\ket{\uparrow\downarrow}+\ket{\downarrow \uparrow })$.
The time evolution of this state under the full Hamiltonian (Eq.~\ref{full_Hamil}) along with the time-dependent term (Eq.~\ref{control}), is shown in Fig.~\ref{fig:population_fidelity}(a). During the evolution, the inverse-Motzkin population $\ket{\downarrow \uparrow }$ is adiabatically suppressed, producing the effective Motzkin state $\ket{\psi_{\mathrm{fin}}}=0.65\ket{00}+0.76\ket{\uparrow\downarrow}+0.02\ket{\downarrow \uparrow }$.

 We measure the performance of the adiabatic-control model using the state fidelity, defined as 
$ 
\mathcal{F}=|\langle \psi (t) | \psi_{\rm Motzkin} \rangle|^2
$,
where $\ket{\psi(t)}$ denotes the instantaneous many-body state of the Rydberg Hamiltonian and $\ket{\psi_{\rm Motzkin}}$ the Motzkin ground state. To experimentally measure the fidelity $\mathcal{F}$, one requires reconstructing the relevant components of the many-body wavefunction $\ket{\psi(t)}$, which can be accessed via state-selective field ionization (SFI) \cite{lu2024probing}. This technique adiabatically ramps an external electric field to ionize different Rydberg states at distinct thresholds; time-resolved signals enable state-selective detection, and repeated runs yield the populations and coherences needed to evaluate $\mathcal{F}$. 
  The corresponding growth in fidelities $\mathcal{F}$ for system sizes $N=2,3,4,6,8$ are presented in Fig.~\ref{fig:population_fidelity}~(b), yielding a final fidelity of $98.84\%$, $97.57\%$, $92.22\%$, $81.14\%$ and $70.06\%$, respectively. 
  The drop in fidelity with increasing system size arises from the rapid growth in the number of unwanted states that requires to be suppressed, which scales approximately as Motzkin number \(M_N \sim 3^N/N^{3/2}\). 
  The growing number of constraints increasingly limits the effectiveness of our protocol at larger \(N\), though it may be mitigated by advanced optimal control techniques.

To validate the entanglement characteristics for our effective Motzkin state, we examine the entanglement entropy scaling and the structure of the reduced density marix.
    \begin{figure}
        \centering
        \includegraphics[width=0.95\linewidth]{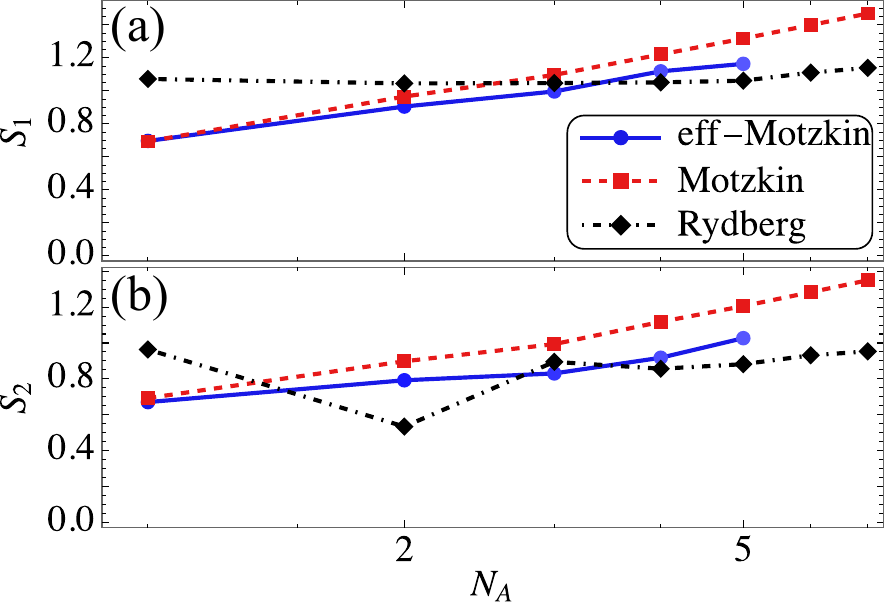}
        \caption{System size scaling of entanglement entropy (a) von Neuman entropy ($S_1$) and (b) Rényi entropy of order 2 ($S_2$), of effective Motzkin ground state (blue-solid), ideal Motzkin ground state (red-dashed) and   Rydberg ground state (black-dot-dashed), for subsystem sizes $N_A=N/2$. The shade of the blue markers encodes the fidelity of the effective Motzkin state, with darker shades corresponding to higher fidelity.}
        \label{fig:entropy_scaling}
    \end{figure}
In Figure~\ref{fig:entropy_scaling}, we consider two types of entanglement entropy measures: 
(i) the von Neumann entropy $
S_{1}(\rho_{A}) 
= -\mathrm{Tr}\!\left[\rho_{A}\log\rho_{A}\right],
$
and (ii) the Rényi entropy of order two
$
S_{2}(\rho_{A})
= -\log\!\left(\mathrm{Tr}\,\rho_{A}^{2}\right),
$
evaluated for a bipartition of size \(N_{A}=N/2\). Here, $\rho_A=\rm{Tr}_B \rho$ is the reduced density matrix of subsystem $A$.
We present the resulting entanglement scaling $S_1(\rho_{A})$ and $S_2(\rho_{A})$, in (a) and (b), respectively of the effective Motzkin ground state (blue-circles), benchmarked against the ideal Motzkin ground state (red-square).  
Both states indicate that they violate the area law and exhibit comparable entanglement scaling across system sizes. Our analysis is restricted to small system sizes, as the exponential growth of the Hilbert space, scaling as $3^N$, renders exact diagonalization computationally prohibitive for larger systems.
For comparison, we also plot the entanglement of the   ground state of the Rydberg Hamiltonian following Eq.~\ref{full_Hamil} (black-diamond), which does not display a consistent scaling trend with system size.  

    \begin{figure}[tb]
        \centering
        \includegraphics[width=0.99\linewidth]{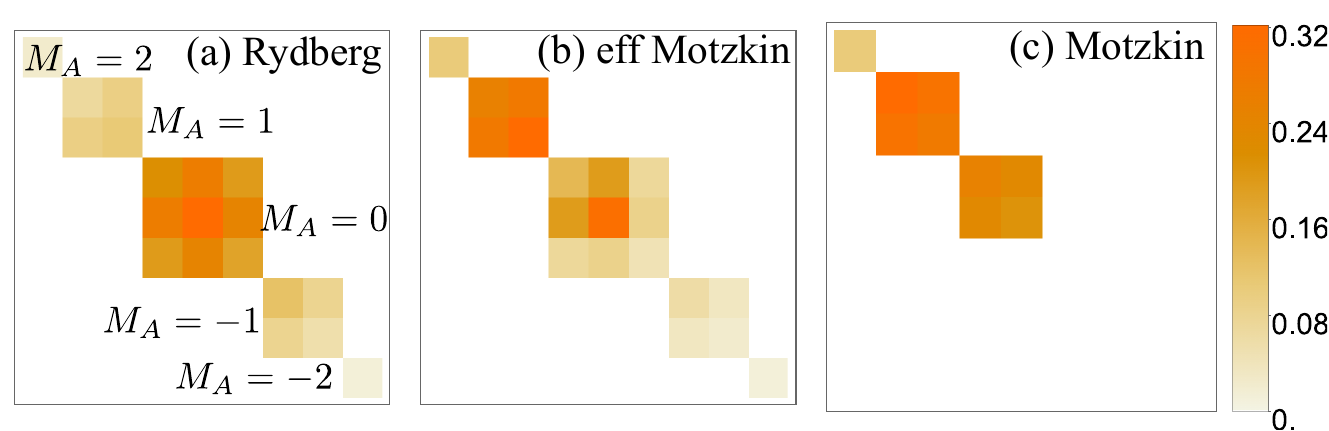}
        \caption{Reduced density matrix structure for system size $N=4$ and sub-system size $N_A=2$ in (a)   Rydberg ground state, (b) effective Motzkin ground state in Rydberg array and (c) Motzkin ground state.The structure consists of five magnetization sector or symmetry blocks $M_A\in \{2,1,0,-1,-2\}$ of the reduced two atom basis $M_A=\sum_{n=1}^2 \langle S_{z,n} \rangle$, where $S_{z} \ket{\uparrow}=+1 \ket{\uparrow}$, $S_{z} \ket{0}=0$ and $S_{z} \ket{\downarrow}=-1 \ket{\downarrow}$.}
        \label{fig:rDM}
    \end{figure}
Beyond entanglement entropy, the structure of the reduced density matrix (rDM) provides a symmetry based understanding of possible deviations in system-size scaling  between different R\'enyi measures \cite{Sugino:2018,Barghathi:2026}.
In the presence of global constraints, the reduced density matrix often displays a block-diagonal structure. Here, this structure is due to the vanishing of the total magnetization $M =\sum S_{z,i} $ of the ground state, where $M \vert\psi_{GS}\rangle = 0$. Thus, we can identify each block based on the local value of $M_A=\sum_{ i \in A} \langle S_{z,i} \rangle$ within subsystem $A$. Using this behavior as a benchmark, Fig.~\ref{fig:rDM} presents a comparative analysis of the rDM for three cases: (a) the   Rydberg ground state, (b) the effective Motzkin ground state realized in the Rydberg platform, and (c) the ideal Motzkin ground state. The rDM of the Rydberg ground state, shown in panel~(a), exhibits a block-diagonal pattern but distributes its weight across all symmetry blocks.  After the application of our adiabatic-control protocol, the resulting effective Motzkin ground state in panel~(b) develops a characteristic redistribution of weight, with an enhancement in blocks corresponding to $M_A\ge 0$.  This pattern closely resembles the rDM from the ideal Motzkin ground state shown in panel~(c), which disallows contributions with $M_A<0$. This analysis provides additional evidence that our engineered state faithfully captures the defining features of the Motzkin ground state.


We now briefly discuss the experimental feasibility of our protocol in realistic Rydberg platforms. A successful implementation of our protocol relies on two key factors: (i) high-fidelity state preparation and (ii) generation of the effective Motzkin state within the coherence time of the system. For the experimental parameters considered here (see Sec.~\ref{SI:interaction_para} in SM~\cite{SI}), we consider $^{87}$Rb atoms and Rydberg states $\ket{81S_{1/2}}$, $\ket{80P_{1/2}}$, and $\ket{80S_{1/2}}$ with nearest-neighbor spacing $r=7\mu$m. In this regime, state preparation can be affected by calibration errors in microwave couplings at the percent level~\cite{ebadi2021quantum,bluvstein2022quantum}, as well as by trap-position fluctuations on the order of tens of nanometers~\cite{browaeys2020many}, which can lead to small variations in the interaction strengths. Nevertheless, high-fidelity many-body states are routinely prepared in current Rydberg-array experiments~\cite{lu2024probing,jiao2025single,chen2024quantum,ebadi2021quantum,bluvstein2022quantum,scholl2021quantum}. The generation of an effective Motzkin state within the coherence time of the system is set by the finite Rydberg-state lifetime, which imposes an upper limit on the protocol timescale. The Rydberg states around $n\sim 80$ exhibit a lifetime of $\tau_{80S}\approx 620~\mu\mathrm{s}$ in the zero-temperature limit~\cite{beterov2009quasiclassical}, leading to an effective coherence time of $\tau_{\mathrm{eff}}\sim \tau/N$ for $N$ atoms. Since the duration of our protocol is typically $20-30~\mu\mathrm{s}$, with $10~\mu$s for ground state preparation and $10-20~\mu$s to the adiabatic control stage, 
decay-induced errors remain at the level of only a few percent for system sizes up to $N\sim 30$, well within the capabilities of current experiments. 

    \ssection{Conclusion and outlook} 
Despite the remarkable properties and potential relevance to quantum technologies, the Motzkin state has largely remained a mathematical construct. Our results demonstrate that the protocol presented here allows a practical realization of the Motizkin state. We further confirm the expected entanglement behaviour of the effective Motzkin state obtained from our protocol.  Our proof-of-principle demonstration of Motzkin spins can be extended to colored Motzkin spins. 
However, the curse of dimensionality can compromise the fidelity of the effective Motzkin state even at smaller system sizes. Nevertheless, this challenge presents a promising avenue for quantum optimal control techniques to enhance state preparation.
 In future works, targeting a given rDM block structure resulting from desired symmetries enables a route towards engineering designer Hamiltonians bypassing the need to optimize state-based fidelity.
 Our work thus opens a pathway toward the controlled preparation of highly entangled ground states in experimentally accessible systems, thereby enriching the available entanglement resources. 

    \ssection{Acknowledgments}
KM and RM acknowledge support from the U.S. National Institute of Standards and Technology (NIST) through the CIPP program under Award No. 60NANB24D218. RM also acknowledge support from the U.S. National Science Foundation (NSF) through the NSF TIP program under Award No. 2534232. This research was partially supported by the National Science Foundation Materials Research Science and Engineering Center program through the UT Knoxville Center for Advanced Materials and Manufacturing (DMR-2309083).
%


\clearpage
\appendix

\section*{Supplemental Material}

\section*{Abstract}

This Supplemental Information provides an overview of the mathematical description of the Motzkin spin model, comparison of Motzkin and Rydberg Hamiltonian, set of realistic interaction parameters used in numerical simulations, and the ground-state preparation protocol.

\section{Motzkin Spin Model}
\label{SI:Motzkin_derivation}

\begin{figure*}[htb]
    \centering
    \includegraphics[width=0.65\linewidth]{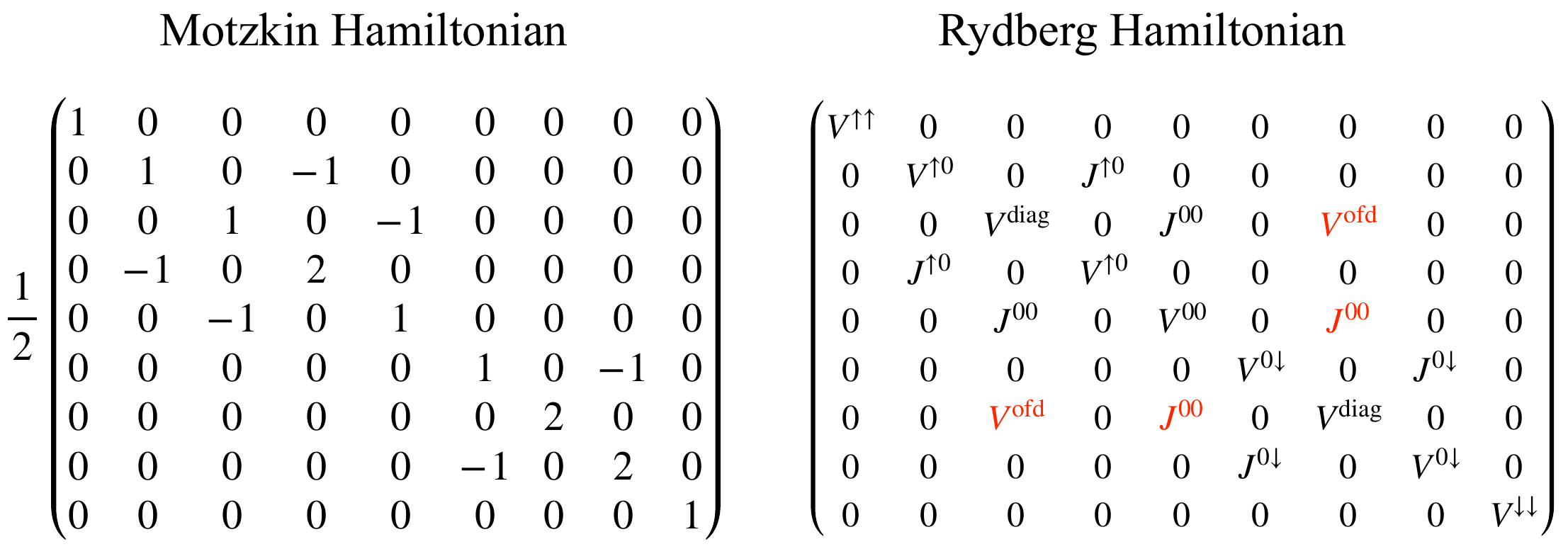}
    \caption{Direct comparison between the Motzkin and Rydberg Hamiltonians. Red-highlighted terms denote contributions not present in the Motzkin Hamiltonian.}
    \label{fig:hamil_matrix}
\end{figure*}

We consider a one-dimensional chain of $N$ spin-1 sites, with a local Hilbert space spanned by three states
\begin{equation}
\{ \ket{u}, \ket{f}, \ket{d} \},
\end{equation}
which encode the three allowed steps of a Motzkin path: \textit{up}, \textit{flat}, and \textit{down}, respectively. In the main text, these states are represented as $\{ \ket{\uparrow}, \ket{0}, \ket{\downarrow} \}$ to streamline notations for both Motzkin spin model and Rydberg-atom quantum simulator. 

A Motzkin path of length $N$ is a sequence
\begin{equation}
\ket{\sigma} = \ket{\sigma_1 \sigma_2 \cdots \sigma_N}, \quad \sigma_i \in \{u,f,d\},
\end{equation}
subject to the constraints:
(i) the path starts at height zero,
(ii) the path ends at height zero, and
(iii) the height never becomes negative at any intermediate site.

These constraints ensure that each up-step is matched by a corresponding down-step. 

The Motzkin ground state is defined as the equal-weight superposition of all valid Motzkin paths,
\begin{equation}
\ket{\psi_{\mathrm{Motzkin}}} =
\frac{1}{\sqrt{\mathcal{N}}}
\sum_{\sigma \in \text{Motzkin paths}}
\ket{\sigma},
\end{equation}
where $\mathcal{N}$ is a normalization constant.

The mathematical construction of the Motzkin Hamiltonian ensuring that the equal superposition of all valid Motzkin paths forms a unique zero-energy eigenstate, relies on the systematic cancellation of pairs of configurations that differ only by the exchange of neighboring steps. For example, for every configuration containing a local segment $\ket{\cdots u f \cdots}$, there exists a corresponding configuration $\ket{\cdots f u \cdots}$ in the Motzkin ground state $\ket{\psi_{\mathrm{Motzkin}}}$. Enforcing destructive interference between such pairs naturally leads to the projector  $\hat{U} = \frac{1}{2}(\rm \ket{ uf} - \ket{fu})(\bra{\rm uf} - \bra{fu})$. In a similar fashion, the pairs $\ket{\cdots d f \cdots}$ and $\ket{\cdots f d \cdots}$, as well as $\ket{\cdots u d \cdots}$ and $\ket{\cdots f f \cdots}$, give rise to the projectors
\begin{eqnarray}
    \hat{U} = \frac{1}{2}(\rm \ket{ uf} - \ket{fu})(\bra{\rm uf} - \bra{fu}), \\
    \hat{D} = \frac{1}{2}(\rm \ket{ df} - \ket{fd})(\bra{\rm df} - \bra{fd}), \\
    \hat{F} = \frac{1}{2}(\rm \ket{ ud} - \ket{ff})(\bra{\rm ud} - \bra{ff}).
\end{eqnarray}

Together, these set of local projectors (S4-S6) enforce zero-energy eigenstate to the the allowed Motzkin path transitions while penalizing configurations that violate the Motzkin constraints. The full Motzkin Hamiltonian can therefore be written as
\begin{eqnarray}
    \hat{H} &=& \hat{U}+\hat{D}+\hat{F} + \ket{d}_1\bra{d} + \ket{u}_N\bra{u}\nonumber \\
    &=& 1/2(|\rm uf \rangle \langle uf| + |\rm fu \rangle \langle fu| - (|\rm uf \rangle \langle fu|+h.c.) \nonumber \\
    & &+|\rm df \rangle \langle df| + |\rm fd \rangle \langle fd| - (|\rm df \rangle \langle fd|+h.c.)
    \nonumber \\
    & & +|\rm ud \rangle \langle ud| + |\rm ff \rangle \langle ff| - (|\rm ud \rangle \langle ff|+h.c.))
\end{eqnarray}

Finally, adopting the encoding $\{ u \rightarrow \uparrow, f \rightarrow 0, d \rightarrow \downarrow \}$ yields the Motzkin Hamiltonian in Eq.~\ref{Motzkin_Hamil} used in the main text,
\begin{equation}
\hat{H}_{\mathrm{Motzkin}} =
\frac{1}2\sum_{i=1}^{N-1} \hat{\Pi}_{i,i+1} + \hat{\Pi}_{\mathrm{boundary}} .
\end{equation}
where the projectors $\Pi_{i,i+1}$ act on $i$th and $(i+1)$-th site of the system as,
\begin{align}
    \Pi_{i,i+1} &= 
   \bigg\{ \ketbra{\uparrow0}{\uparrow0} + \ketbra{0\uparrow}{0\uparrow} +
    \ketbra{\downarrow0}{\downarrow0} \nonumber \\
    &+ \ketbra{0\downarrow}{0\downarrow} +  \ketbra{00}{00}\bigg\} +
   \bigg[ \ketbra{\uparrow\downarrow}{\uparrow\downarrow} \bigg]  \nonumber \\
    &- \bigg(\ketbra{\downarrow0}{0\downarrow} +\ketbra{\uparrow0}{0\uparrow} + \ketbra{\uparrow\downarrow}{00} + \text{h.c.}\bigg) ,
    \label{Motzkin_Hamil_SI}
\end{align}

The boundary term penalizes configurations that violate the Motzkin path constraints at the edges,
\begin{equation}
\hat{\Pi}_{\mathrm{boundary}} =
\ket{\downarrow}_1\!\bra{\downarrow} + \ket{\uparrow}_N\!\bra{\uparrow} .
\end{equation}

We present a direct comparison between the Motzkin Hamiltonian and the Rydberg Hamiltonian, defined in Eq. 5 of the main text, in Fig.~\ref{fig:hamil_matrix}, where the matrix form is in the basis
\begin{equation}
\{ \ket{\uparrow \uparrow }, \ket{ \uparrow 0}, \ket{\uparrow\downarrow}, \ket{0 \uparrow }, \ket{00}, \ket{0\downarrow}, \ket{\downarrow \uparrow }, \ket{\downarrow 0}, \ket{\downarrow \downarrow } \},
\end{equation}
where the red-highlighted terms denote contributions not present in the Motzkin Hamiltonian.

\section{Realistic Parameters for modeling Motzkin spins}
\label{SI:interaction_para}

In this section, we provide the interaction parameters used to model the effective Motzkin spin system in a realistic Rydberg-atom platform. These parameters are obtained by calculating the dipole-dipole, van der Waals interaction strengths and F\"orster resonance between relevant Rydberg states using the \texttt{Pairinteraction} package \cite{weber2017calculation}. The calculations are performed for the atomic states and geometries considered in our protocol, enabling an accurate description of the effective Hamiltonian. Throughout this section, positive (negative) interaction coefficients correspond to repulsive (attractive) interactions.

\subsection{Rubidium Parameters (Adiabatic-control model)}

For $^{87}$Rb atoms:
\begin{align}
\ket{\uparrow} &= \ket{81S_{1/2}, m_j=+1/2}, \\
\ket{0} &= \ket{80P_{1/2}, m_j=+1/2}, \\
\ket{\downarrow} &= \ket{80S_{1/2}, m_j=+1/2}.
\end{align}

Nearest-neighbor spacing is $r = 7~\mu$m and the quantization-axis angle is $\theta = 35.1^\circ$. Table~\ref{table:Rb} lists the dipole-dipole ($J^{\alpha\beta}$) and van der Waals ($V^{\alpha\beta}$) interaction 
coefficients used in our numerical simulations with adiabatic-control model.

\begin{table}[htb]
\centering

\begin{tabular}{|c |c |c|}
\hline
Interaction type &Interaction & Strength (GHz) \\
\hline
&$J^{\uparrow0}$  & $-26.363$ \\
Dipole-dipole &$J^{\downarrow0}$  & $-28.711$ \\
($\mu m^3$) &$J^{00}$  & $-27.512$ \\
\hline
&$V^{\uparrow0}$  & $3883.332$ \\
&$V^{\downarrow0}$  & $2766.837$ \\
van der Waals &$V^{\uparrow\uparrow}$  & $4648.377$ \\
($\mu m^6$)&$V^{\downarrow\downarrow}$  & $3878.318$ \\
 &$V^{00}$  & $1484.518$ \\
\hline
\end{tabular}
\caption{Dipole-dipole (in units GHz $\mu m^3$)  and van der Waals interaction (in units GHz $\mu m^6$) coefficients for Rydberg states of $^{87}Rb$ atoms.}
\label{table:Rb}
\end{table}

\begin{figure}[htb]
    \centering
    \includegraphics[width=0.95\linewidth]{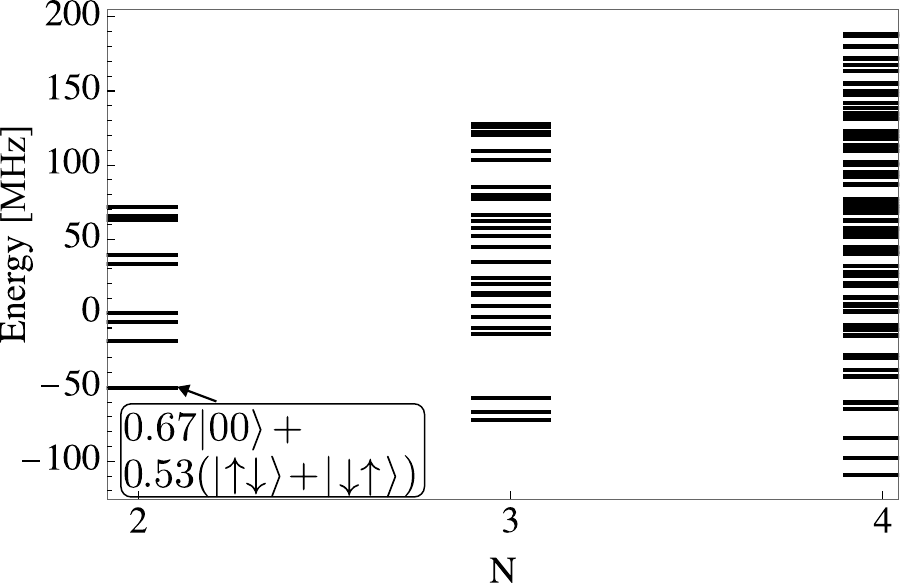}
    \caption{Energy levels of the 1D spin-1 Rydberg system with $^{87}$Rb for $N=2,3$ and $4$. }
    \label{fig:energy_levels}
\end{figure}

We present the energy spectra of the 1D Rydberg chain with above set of parameters for system sizes 
$N = 2,3,$ and $4$. To obtain the spectra, we diagonalize the full Rydberg Hamiltonian for open boundary conditions. Fig.~\ref{fig:energy_levels} shows the numerically obtained eigenenergies for each system size. For all system sizes $N=2,3,4$, the spectrum remains gapped, with a well-separated ground state.

For reference, we highlight the ground-state of the Rydberg Hamiltonian for $N=2$, $\ket\psi_{GS}=0.67\ket{00} + 0.53(\ket{\uparrow\downarrow}+ \ket{\downarrow \uparrow })$ with energy $-50.3$ MHz.
 For $N=3$ and $4$, the corresponding effective Motzkin ground state 
     has energies $-72.3$ MHz and $-109.1$ MHz, respectively.

\subsection{Cesium Parameters (Fine-tuning Model)}


\begin{table}[htb]
\centering

\begin{tabular}{|c |c |c|}
\hline
Interaction type &Interaction & Strength (GHz) \\
\hline
&$J^{\uparrow0}$  & $13.369$ \\
Dipole-dipole &$J^{\downarrow0}$  & $13.966$ \\
($\mu m^3$) &$J^{00}$  & $16.763$ \\
\hline
&$V^{\uparrow0}$  & $621.365$ \\
&$V^{\downarrow0}$  & $480.496$ \\
van der Waals &$V^{\uparrow\uparrow}$  & $3157.912 $ \\
($\mu m^6$) &$V^{\downarrow\downarrow}$  & $3656.713$ \\
&$V^{00}$  & $-437.895$ \\
\hline
F{\"o}rster& $V^{\rm diag/ofd}$ &$420.38$ \\
\hline
\end{tabular}
\caption{Dipole-dipole (in units GHz $\mu m^3$) and van der Waals (in units GHz $\mu m^6$) interaction coefficients for Rydberg states of $^{133}Cs$ atoms.}
\label{table:Cs}
\end{table}

For $^{133}$Cs atoms:
\begin{align}
\ket{\uparrow} &= \ket{81S_{1/2}, m_j=+1/2}, \\
\ket{0} &= \ket{80P_{3/2}, m_j=+3/2}, \\
\ket{\downarrow} &= \ket{80S_{1/2}, m_j=+1/2}.
\end{align}
 Nearest-neighbor separation \(r = 8.69~\mu\mathrm{m}\), F{\"o}rster off-resonant detuning \(\Delta=-223.2\)MHz and quantization-axis angle is chosen to be \(\theta = 9.376^\circ\).
Table~\ref{table:Cs} lists the dipole-dipole ($J^{\alpha\beta}$) and van der Waals ($V^{\alpha\beta}$) interaction 
coefficients used in fine-tuning model.

\begin{figure}[htb]
    \centering
    \includegraphics[width=0.95\linewidth]{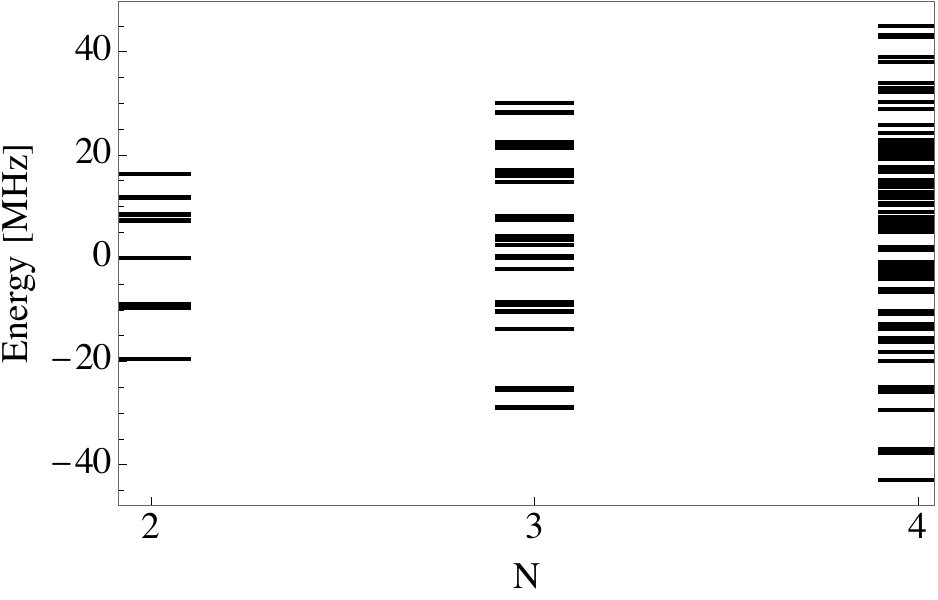}
    \caption{Energy levels of the 1D spin-1 Rydberg system with $^{133}$Cs for $N=2,3$ and $4$. }
    \label{fig:energy_levels_Cs}
\end{figure}

The energy spectra of the 1D Rydberg chain with above set of parameters for system sizes 
$N = 2,3,$ and $4$ is shown in Fig.~\ref{fig:energy_levels_Cs}.

    \begin{figure}[htb]
        \centering
        \includegraphics[width=0.99\linewidth]{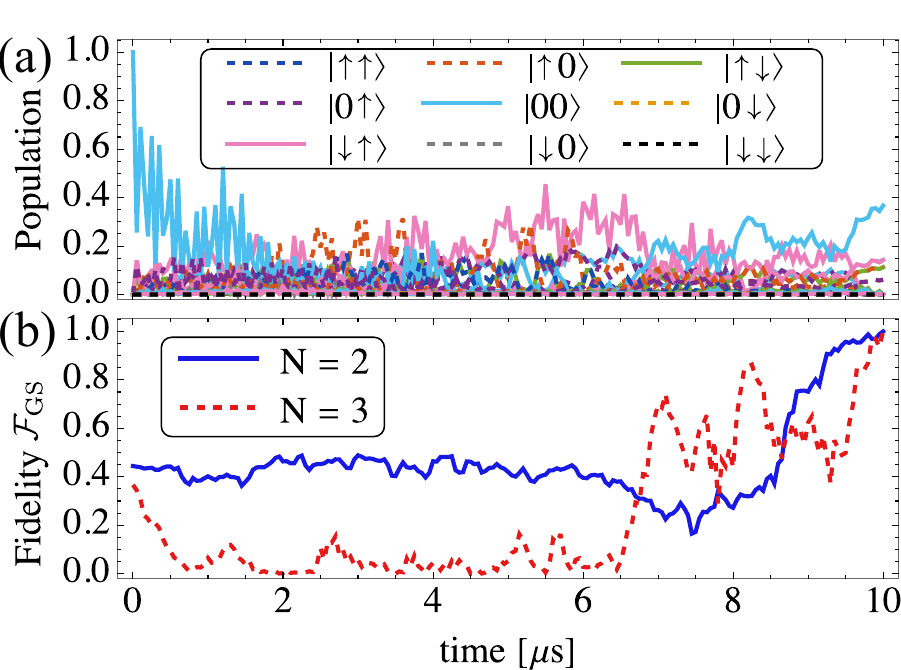}
        \caption{Ground state preparation in the Rydberg system using quantum optimal control with a couple of microwaves. (a) Population dynamics for $N=2$ with each line indicating a many-body state marked in the legends. (b) Fidelity $\mathcal{F}_{\rm GS}=|\langle \psi (t) | \psi_{\rm GS} \rangle|^2$ with respect to Rydberg ground state for $N=2$ and $N=3$.}
        \label{fig:state_prep}
    \end{figure}
    
\section{Ground-State Preparation Protocol}
\label{SI:ground_state_prep}

To prepare the Rydberg ground state, we consider a ground-state preparation protocol in which the system is initialized in the product state $ \ket{\psi_{\mathrm{ini}}} = \ket{00\cdots 0}$, and subsequently evolved toward the target ground state using time-dependent fields. In Rydberg atoms, energy levels are typically separated by frequencies in the GHz regime, and transitions between states with angular momentum difference $\Delta l=\pm 1$ are dipole-allowed and can be driven using microwave fields. Here, we consider two microwave fields 
that resonantly drive the transitions
$\ket{\uparrow}\leftrightarrow\ket{0}$ and $\ket{\downarrow}\leftrightarrow\ket{0}$, which are described by
\begin{align}
    \hat{H}_{\mathrm{mw}}(t) = &\sum_{i=1}^N \sum_{\alpha\in\{\uparrow,\downarrow\}} \bigg[\frac{\bar{\Omega}_i^{\alpha }(t)}{2}(\ket{0}_i\bra{\alpha}+h.c.)\nonumber\\
    &+
    \bar{\delta}_i^{\alpha}(t)\,\ket{\alpha}_i\bra{\alpha}\bigg],
    \label{microwave}
\end{align}
where $i$ labels the atomic site, $\bar{\Omega}_i^{\alpha }$ and $\bar{\delta}_i^\alpha(t)$ denotes the Rabi frequency and detuning, respectively, of the microwave coupling $\ket{\alpha}\leftrightarrow\ket{0}$, where $\alpha\in\{\uparrow,\downarrow\}$.
Subsequently, Quantum optimal control (GRAPE) \cite{khaneja2005optimal,de2011second}  is then used to shape the pulses, steering the system into the Rydberg ground state. The same fields can then be used for adiabatic control.

In Fig.~\ref{fig:state_prep}~(a), we show the time dynamics of state populations for $N=2$, demonstrating the evolution of $\ket{\psi_{\rm ini}}=\ket{00}$ into gradual mixing 
    of different states, finally arriving at the Rydberg ground state $\ket{\psi_{\rm GS}}=0.67\ket{00}+0.53(\ket{\uparrow\downarrow}+\ket{\downarrow \uparrow })$. We further show the fidelity $\mathcal{F}_{\rm GS}=|\langle \psi (t) | \psi_{\rm GS} \rangle|^2
$ with respect to the Rydberg ground state for $N=2$ and $N=3$ in Fig.~\ref{fig:state_prep}~(b). 
    The fidelity increases over time and saturates near its target value of $\mathcal{F}_{\rm GS}=0.9999$ in protocol preparation time 
    $T_{\rm GS}= 10~\mu$s.
This state constitutes the initial condition for the adiabatic preparation into the Motzkin ground state 
described in the main text.
    
\end{document}